\documentclass{article}

\usepackage{graphicx} 
\usepackage[a4paper]{geometry}
\usepackage[switch]{lineno}
\usepackage[justification=centering]{caption}
\usepackage{algorithm}
\usepackage[noend]{algpseudocode}
\usepackage{longtable}
\usepackage{siunitx}
\usepackage{bm}
\usepackage{color}
\RequirePackage[bookmarks, bookmarksopen=true, plainpages=false, pdfpagelabels, pdfpagelayout=SinglePage, breaklinks = true]{hyperref}
\usepackage[title]{appendix}
\usepackage{caption}

\algblockdefx{ForAllP}{EndFAP}[1]%
{\textbf{for all }#1 \textbf{do in parallel}}%
{\textbf{end for}}

\algblockdefx{Critical}{EndCritical}[1]%
{\textbf{critical }#1}%
{\textbf{end critical}}

\title{Divide-and-Conquer Strategy for Large-Scale Eulerian Solvent Excluded Surface} 

\author{Rundong Zhao$^1$, Menglun Wang$^2$, Yiying Tong$^1$\footnote{
		Corresponding author:		Email: ytong@msu.edu}, Guo-Wei Wei$^{2,3,4}$\footnote{
		Corresponding author.		Email: wei@math.msu.edu} \\
$^1$ Department of Computer Science and Engineering,\\
Michigan State University, MI 48824, USA.\\
$^2$ Department of Mathematics, \\
Michigan State University, MI 48824, USA.\\
$^3$ Department of Electrical and Computer Engineering,\\
Michigan State University, MI 48824, USA. \\
$^4$ Department of Biochemistry and Molecular Biology,\\
Michigan State University, MI 48824, USA. \\
}
\date{\today} 

\usepackage{color}

\begin{document}

\maketitle 
\pagenumbering{roman}

	
\begin{abstract}
\noindent {\bf Motivation:}
Surface generation and visualization are some of the most important tasks in biomolecular modeling and computation.   Eulerian solvent excluded surface (ESES) software provides analytical  solvent excluded surface (SES) in the Cartesian grid, which is necessary for simulating many biomolecular electrostatic and ion channel models. However, large biomolecules and/or fine grid resolutions give rise to excessively large memory requirements in ESES construction.   We introduce an out-of-core and parallel algorithm to improve the  ESES  software.\\
{\bf Results:} The present approach drastically improves the spatial and temporal efficiency of ESES. The memory footprint and time complexity are analyzed and empirically verified through extensive tests with a large collection of  biomolecule examples. Our results show that our algorithm can successfully reduce memory footprint through a straightforward divide-and-conquer strategy to perform the calculation of arbitrarily large proteins on a typical commodity personal computer. On multi-core computers or clusters, our algorithm can reduce the execution time by parallelizing most of the calculation as disjoint subproblems. Various comparisons with the state-of-the-art Cartesian grid based SES calculation were done to validate the present method and show the improved efficiency. This approach makes ESES a robust software for the construction of analytical solvent excluded surfaces.\\
{\bf Availability and implementation:} \url{http://weilab.math.msu.edu/ESES}.
\end{abstract}

 \renewcommand{\thepage}{{\arabic{page}}}
%
\section{Introduction}
As a principal tool to study the biomolecular world, molecular modeling and analysis have an increasing impact in computational biology. The accuracy and efficiency of molecular modeling and analysis are often crucial in enabling more sophisticated downstream research. Researchers have made persistent efforts in reconstructing and visualizing the details of biomolecules through various simplifications, including the ball-and-stick model by von Hofmann, dated back to 1865, and the ribbon diagram by Richardson for illustrating protein structures. However, in order to simulate physical phenomena like the electrostatic distribution of macromolecules in a cellular environment, a much more elaborate model is needed to describe the interface between solvent and solute regions. The van der Waals surface (i.e., ``atom and bond'' model by Corey and Pauling in 1953) was introduced to describe such interfaces, where each type of atoms was described by a sphere with the corresponding van der Waals radius. For various simulations and geometric smoothness, concepts of solvent accessible surface (SAS) \cite{connolly1983solvent, richmond1984solvent} and solvent excluded surface (SES) \cite{lee1971interpretation, richards1977areas} were built on top of the van der Waals radii. SAS captures the trajectory of the center of a probe atom rolling on the van der Waals surface as the interface delineating the boundary of regions accessible by the center of any solvent molecule. SES is defined by the boundary of the union of all possible outside probe balls, and thus consists of three types of patches. Specifically, convex patches, where the probe touches one of the atoms of the molecule, saddle patches, where the probe touches two atoms, and concave patches, where the probe touches three or more atoms, are parts of an SES for a biomolecule.

All of these models still fail to guarantee the interface smoothness, as singularities and sharp edges cannot be completely avoided for the aforementioned geometry models for biomolecules. Minimal molecular surface (MMS) based on the mean curvature flow was introduced to resolve this issue \cite{bates2008minimal, Bates:2009}. Various Gaussian surfaces \cite{Blinn:1982,Duncan:1993, QZheng:2012,ZYu:2008, MXChen:2011,Decherchi:2013}, skinning  surface \cite{HCheng:2009} and flexibility-rigidity index (FRI) surface \cite{KLXia:2013d, Opron:2014} have been proposed to achieve a similar goal. Another limitation for these models is that they only reflect the static or instantaneous shape in vacuum. In practice, solvent and solute interactions, making a static interface inaccurate for certain biophysical analysis.    Thus, various solvent-solute interactive boundaries were proposed \cite{dzubiella2006coupling, wei2010differential}. However, despite its weaknesses, SES remains the most favorable model among biophysicists, due to its simplicity and effectiveness in capturing the interface of solvent and solute through its definition, with which various physical phenomena can be described with a reasonable accuracy.

Many software packages were developed to calculate SES \cite{Rocchia:2002}.  Among them, MSMS is of considerable influence \cite{sanner1996reduced}. Built on top of MSMS, there are various software packages for different purposes. For the Lagrangian representation, a triangle mesh can be directly constructed for the three different types of patches followed by a concatenation. Nevertheless, MSMS is known for its efficiency and robustness issues, which often occur when large protein molecules and fine resolutions are required \cite{JCC:JCC24682}. Moreover, many biophysical phenomena are happening not only on the surface, but inside the encapsulated volume of the molecules. To address these issues and meet the requirements of volumetric output, Liu {\it et al.}~\cite{JCC:JCC24682} introduced an Eulerian solvent excluded surface (ESES) approach as an alternative for surfaces represented as intersections and normals with a regular Cartesian grid. The ESES algorithm starts with a list of atoms describing the molecule enclosed by a regular Cartesian grid. Based on the three different types of patches for SES, all grid points are classified as either inside or outside with respect to SES. Finally, intersection points are computed on each mesh line with two ends on opposite side of the interface. It is also straightforward to be converted into the Lagrangian representation, i.e., a triangle mesh, through the marching cubes algorithm. Although high accuracy and robustness are well addressed with this method, it often suffers from  the lack of efficiency as well as overly large memory requirements, as a full regular grid has to be maintained. The ESES algorithm is sequential, which results in long execution time especially when the grid resolution increases due to a fine grid spacing or large protein complexes  with many atoms \cite{li2016multiscale}.

In this work, we propose an out-of-core parallelizable version of ESES, in which we divide the bounding box of the molecule into tiled sub-blocks based on the localized nature of the problem. By performing the computation based on local information, one can avoid keeping the whole grid and all the atoms in memory, and at the same time, distribute the computation to multiple processors. Thus, for large molecules or fine grids, both space and time efficiency can be substantially improved. By restricting the active subdomains that are being executed, the whole procedure can always be done on a personal computer (PC) with a fixed memory, e.g., 2GB. Testing and comparison are done on the 2016 core set of PDBbind database (http://www.pdbbind.org.cn/), with additional validation by users worldwide through the authors' website for the project.

The rest of the paper is organized as follows. Section \ref{Sec:Algorithm} discusses design of the improved algorithm with locality. Section \ref{sec:Spatial} is devoted to the space and time complexity analysis of the proposed algorithm. The validation and comparison of our results are carried out in  Section \ref{Sec:Validation}. Section \ref{Sec:Conclusion} concludes the paper with a comment on future work.

\section{Algorithm}\label{Sec:Algorithm}
\subsection{Recap of ESES}
As we aim at improving the efficiency of ESES \cite{JCC:JCC24682}, we assume the same input, a list $\mathcal{A}$ of atoms, represented by the center location $\bm{c}_i$ and the corresponding van der Waals radius $r_i$ for each atom,
\begin{equation}\label{eq:input}
\mathcal{A}=\{(\bm{c}_i,r_i)\}_{i=1\dots N},
\end{equation}
where $N$ is the number of atoms.

We also assume the same output: first, a 3D array of Boolean indicating whether each grid point is inside the molecule surface,
\begin{equation}\label{eq:output_inside}
\rm{Inside}\left[i,j,k\right]=\left\{
                                \begin{array}{ll}
                                  1, & \hbox{$(ih,jh,kh)\in M$} \\
                                  0, & \hbox{otherwise,}
                                \end{array}
                              \right.
\end{equation}
where $h$ is the grid spacing, and $M$ is the volume enclosed by SES; and second, a set of intersection points between grid edges and the SES
\begin{equation}\label{eq:output_intersection}
\mathcal{I}=\{\bm{s},\bm{t},\lambda_{st} \}_{\bm{st}\in\mathcal{E}},
\end{equation}
where $\bm{s}$ and $\bm{t}$ are grid points {with $\bm{st}$ representing the corresponding edge, and $\mathcal{E}$ is the grid edge set with  two grid points adjacent to each other and ${\rm{Inside}}[\bm{s}]\neq {\rm{Inside}}[\bm{t}]$.} The location of the intersection point can be computed through
\begin{equation}\label{eq:location}
\bm{p}=\lambda_{st} \bm{s} + (1-\lambda_{st})\bm{t}.
\end{equation}
Note that, using the connectivity construction procedure in the standard marching cubes algorithm, we can also output a triangle mesh based on the Eulerian output.

\subsection{Overview}
The main idea for reducing the main memory requirement
is through a simple domain decomposition without the need to explicitly handle the boundary matching problem. Owing to the locality of the ESES algorithm, we can straightforwardly decompose the computational domain into many non-overlapping subdomains. With each subdomain retaining only a small number of atoms that are less than the largest van der Waals radius away from the subdomain boundary, we have all the information necessary to determine the inside and the outside information, as well as the intersections of the SES surface and the grid edges within the subdomain. Thus, we can successfully reduce the memory footprint by controlling the size of each subdomain so as to fit within the main memory limit  of a typical PC. The memory storage for the list of atoms relevant for the subdomain is negligible compared to the storage requirement for the subgrid, since the grid spacing in practical applications would typically be smaller than the van der Waals radius of the smallest atom.

\begin{figure}
	\centering
	\includegraphics[width=0.99\textwidth]{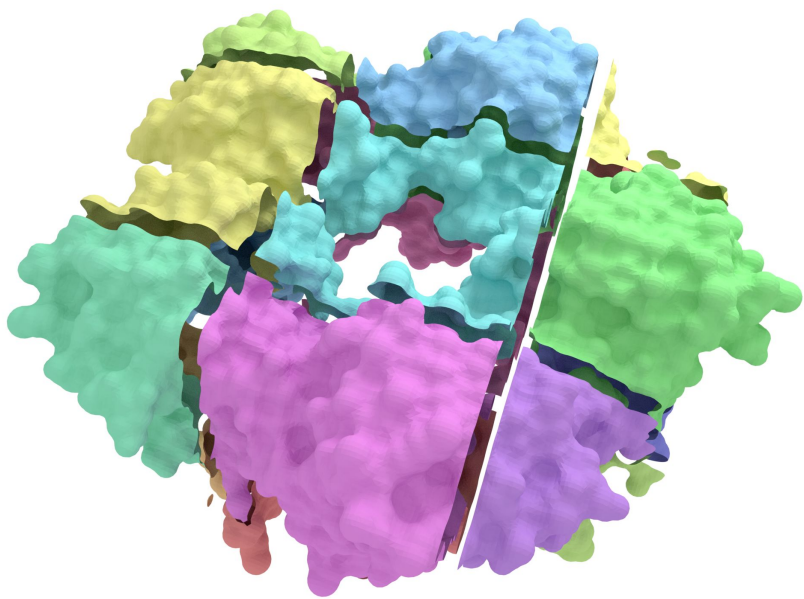}
	\caption{\textbf{Illustration of subdomain based algorithm}. The bounding box of the whole molecular surface is  divided into several non-overlapping subdomains, which can be computed independently with a small memory footprint. Finally, patches of the SES from each subdomain can be assembled into a watertight surface identical to the one constructed with ESES. Note that the patches form a watertight surface, the gaps between the adjacent patches from different subdomains are added for visualization only.}
	\label{Fig:overview}
\end{figure}

As shown in Figure~\ref{Fig:overview}, patches rendered with different color belong to different subdomains, and they can be independently constructed by intersection detection locally within the corresponding subdomain, followed by using the marching cubes algorithm. With a direct concatenation of all the output, we can construct the whole molecular surface. It is also possible for the downstream applications to choose only the subdomains relevant for the calculation that they perform.

When designing a parallelizable out-of-core algorithm, the first and foremost problem to deal with is to analyze the dependence among different steps of the procedure or different parts of the data. In this section, we examine the four main stages of the ESES algorithm \cite{JCC:JCC24682}:
\begin{itemize}
  \item construction of the grid and the analytical expression for patches of the SES,
  \item classification of the grid points to the inside points or the outside points of the SES,
  \item calculation of the intersection between grid edges and the SES,
  \item and assembly of the output.
\end{itemize}

When performed in a subdomain of the entire domain, the first three steps need to performed in the given order, but they do not have data dependence to the calculation done on any other subdomain. For instance, the classification of any grid point can be locally determined by the nearby atoms, more precisely, atoms at a distance less than  the sum of the probe radius and the largest van der Waals radius. Similarly, where the intersection is on a grid edge depends only on the SES analytical patch expressions determined by nearby atoms. Thus we can set up one thread per subdomain, without any possibility of  race conditions, i.e., our output is independent on the timing of each thread.

If file I/O exchanges are done through sequential devices such as a hard drive, the final output step would have to be done after finishing the previous three steps. On the other hand, if the final output is to reside in a random access memory, once the size of output for each subdomain is determined, it is possible to assemble the output in sublinear time. If the file system allows for concatenation without moving data blocks, it is also possible for each thread to write to a different file, and concatenate them in a time linear manner with respect  to the number of subdomains.

As in ESES, for the robustness of the calculation, there are some grid points left undetermined as inside or outside in the second step, and only finalized in the third step after the intersections of nearby grid edges are determined \cite{JCC:JCC24682}. Nevertheless, this procedure will only have a local data dependence. Thus, partitioning the whole grid into several subdomains does not change the final classification of such points. As confirmed by our large set of test results, the uncertain grids are rare as indicated by \cite{JCC:JCC24682}, and when they indeed exist, their classification in the subdomain based approach is identical to that in the original ESES.

In sum, we can safely assume that there is no communication of information between intermediate results from different subdomains. This implies that we can do out-of-core calculation by loading only one subdomain, and/or parallelizing the calculation by simultaneously initiating one thread per subdomain.

\subsection{Decomposition to subdomains}
The entire calculation domain in ESES is a regular Cartesian grid inside a cuboid. Typically, it is constructed as a tight bounding box of the list of the atoms, padded with a few additional layers of grid cells to provide some margin for easy handling of the boundary cells.

Therefore, it is a natural choice to design the subdomain as non-overlapping cubes with the same number of grid cells in each of the three dimensions. The domain can be extended slightly if the size of the original entire domain in any direction is not a multiple of the size of the subdomain. We will call a cubic subdomain as a block, following the similar term used in CUDA parallel thread mapping design.

By focusing on the local computation within each block, the memory footprint is mainly determined by the size of the block, since we only need to keep one block in the main memory at a time. Some memory storage is required to store the list of atoms relevant for the analytical SES patch construction. If we store the entire list of atoms, for large molecules, it can still require a large block of memory, and may require more time to perform the nearest neighbor search. Fortunately, due to the localized nature of the calculation, it is possible to determine whether an atom is inside or within a small distance from a block. The rest of the atoms do not need to reside in the memory, and one can treat the calculation inside the block {as if it were} for a smaller molecule, without any unfavorable effect on the final output.

\subsection{Index mapping}
\begin{figure}
	\centering
	\includegraphics[width=0.7\textwidth]{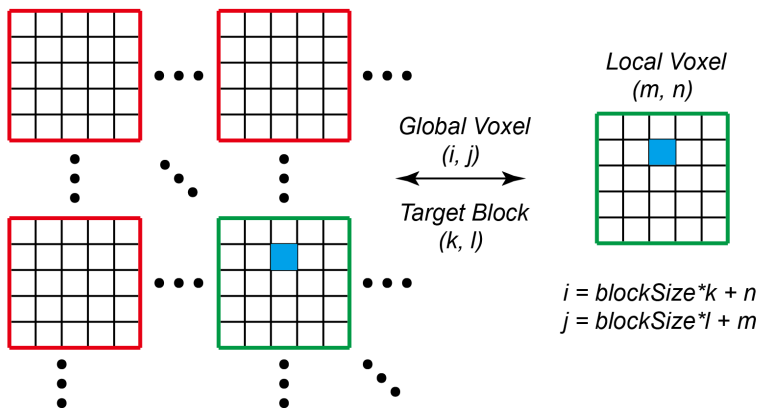}
	\caption{\textbf{2D Index mapping Example.} All the indices are 0-based. Given a cell with global coordinates $(i, j)$ in block $(k, l)$ (green) and its local coordinates $(m, n)$ within the block, the one-to-one mapping is straightforward as illustrated. Note that $(m,n)$ would require fewer bits to store than $(i,j)$. This is typical in parallel processing such as CUDA threads. }
	\label{Fig:index}
\end{figure}

After the computation is done for each individual block, we need to map the result within that block to the original domain. The index mapping is similar to the CUDA parallel thread mapping design for 3D. We illustrate the basic idea in 2D through Figure~\ref{Fig:index}. Each block has coordinates $(b_x,b_y,b_z)$ indicating the position of its top left back corner. Assuming that the grid cell count along one edge of each block is $(b_s)$, for a grid point with local coordinates $(i,j,k)$ in the block, its corresponding global coordinates are found through the following function
\begin{equation}\label{eq:map}
\mathrm{LocalToGlobal}(i,j,k;b_x,b_y,b_z)= (b_s b_x+i, b_s b_y+j, b_s b_z+k).
\end{equation}

\subsection{Subdomain boundary treatment}
While the subdomains are not overlapping with the outer boundary of the total computational domain, they may intersect at a zero-measure set, such as a common rectangle, a common line segment, or a common point. Thus, during the final assembly of the output from all the blocks, the boundary grid points and boundary grid edges are to be carefully handled. Otherwise, redundant information for grid point classification and intersection points on   subdomain boundaries  may appear in the output. For instance, whether a grid point is inside may be duplicated up to 8 times, if it is at the corner point shared by 8 subdomains. Similarly, the intersection information may also be duplicated up to 4 times, if a grid edge is shared by 4 subdomains.

One way to eliminate the redundancy is to add a post-processing step. However, using ideas commonly used for eliminating such redundancies, we can directly avoid the generation of redundant data. This more efficient approach is based on the partitioning of the domain into truly disjoint   subdomains, each of which has the form of a half-open half-closed box, in other words, the Cartesian product of three half-open half-closed intervals:
$$[b_s b_x, b_s b_x+ b_s)\times[b_s b_y, b_s b_y+ b_s)\times[b_s b_z, b_s b_z+ b_s).$$

\begin{figure}
	\centering
	\includegraphics[width=0.7\textwidth]{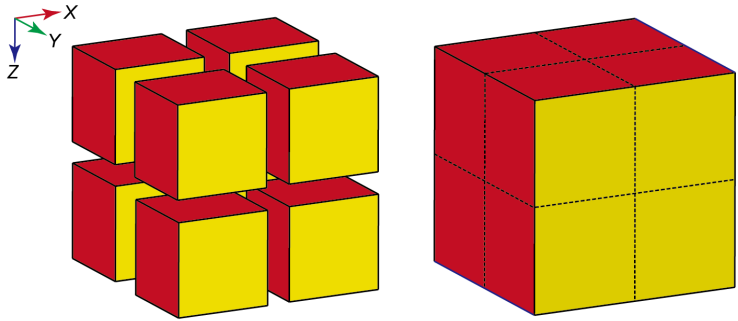}
	\caption{\textbf{Redundancy elimination.} For each subdomain, we only keep the information on faces with a normal along negative axis directions (red), the other faces (yellow) are omitted because they have already been accounted for by adjacent subdomains. After concatenation (dashed line), only the output for faces of the entire grid with normals along the positive axis directions is missing. However, with the margin padded to the molecule when constructing the domain, they do not contain any intersection information to begin with.}
	\label{Fig:redundent}
\end{figure}

Thus, only the grid points and grid edges that lie on the left, top and back faces of the subdomain are considered for the output, while the front, right and bottom faces are ignored as shown in Figure~\ref{Fig:redundent}. When taking the union of the output from the blocks, we eventually omit all the grid points and grid edges that are on the front, right and bottom boundary faces of the entire  domain. Fortunately, by leaving the sufficient margin as mentioned earlier, the whole domain is a bounding box of the molecule, and no inside points or intersection points exist on those faces.

\subsection{Pipeline}
Algorithm~\ref{alg:paraeses} provides the pseudocode of the main procedures for the parallelized version of ESES. The function \textit{ESES} is identical to the original procedure introduced in \cite{JCC:JCC24682}. Each \textit{block} (subdomain) uses a unified data structure to store all the necessary information for the part of ESES computation within that block, including the subgrid, the list of relevant atoms, and the output---inside/outside information for grid points and intersection locations on grid edges.
\begin{algorithm}
	\caption{ParaESES Algorithm}\label{alg:paraeses}
	\begin{algorithmic}[1]	
		\Function{ParaESES}{$data, probeRadius, gridSize, margin$}	
		\State CreateBlocks($data, blocks$)
		\ForAllP {$b \in blocks$}
			\State AssembleRelevantAtoms($b$)
			\State GlobalToLocalIndexMapping($b, bLocal$)
			\State ESES($bLocal, probeRadius, gridSize, margin$)
			\State RemoveDuplicates($bLocal$)
			\State LocalToGlobalIndexMapping($b, bLocal$)
			\Critical
				\State OutputInfo($b$)
			\EndCritical
		\EndFAP
		\EndFunction
	\end{algorithmic}
\end{algorithm}

\section{Spatial and temporal complexity analysis}\label{sec:Spatial}
Our treatment does reduce the total memory requirement but not the total amount of computation. The number of grid point classification operations to perform, and the number of edge and SES intersection tests are not reduced. Nevertheless, as we divide these task into subdomains of the entire domain, we can either handle the previously intractable problem on a commodity  PC with few gigabytes of memory, or greatly reduce the amount of time on a computer cluster or a multi-core computer.

As the number of atoms grows larger or as the resolution becomes finer, the number of grid cells increases asymptotically at $O(whd/h^3)$, where $h$ is the grid spacing, and $w$, $h$, and $d$ are the width, height, and depth of the bounding box of the molecule respectively. When dealing with large proteins in the original ESES~\cite{JCC:JCC24682} on fine grids, the memory storage is essentially cubic to the number of cells along an edge of the box domain. For instance, a grid with the size of $1000\!\times\! 1000\!\times\! 1000$ would require roughly $n$ Giga-Byte if each grid point or grid cell requires $n$ Byte storage. As ESES requires the data to reside in the main memory for the grid point classification and grid edge intersection, it cannot fit in the memory of a regular PC, even with virtual memory.

Our straightforward domain decomposition into blocks can effectively shrink the memory footprint, i.e., the maximum memory requirement at any point of the calculation, since all the information required for the localized calculation is associated with the block (or subdomain). No matter how large the original grid size is, we can treat the problem as if it is for the subgrid of size $b_s^3$, as long as we choose each block to be of size $b_s\!\times\!b_s\!\times\!b_s$.  Specifically, there will be $b_s^3$ grid points to classify, and $3(b_s-1)b_s^2$ grid edges to check for intersections when processing one block. The overhead introduced to handle the boundary of the blocks is negligible, since for each block, there will be $3b_s^2$ duplicate grid points and $6(b_s-1)b_s$ duplicate grid edges to check. So in terms of each grid block, there will be an approximate ratio of $O(1/b_s)$ overhead. As mentioned in the previous section, we introduced a procedure to determine which atoms can influence a particular block, which brings an overhead that is also negligible to the dominating time and space requirement for the grid. This part of the overhead can be further reduced by any spatial data-structure such as a kd-tree, since the construction of the list of atoms within a block is just a query for spatial database entries within a certain spatial range. However, an overly small block size can increase the proportion of the memory the overhead requires, so we do not recommend to aggressively reduce the block size. Fortunately, in practice, even the memory of any modern smart phone can easily accommodate the block size with $b_s=32$.

If multi-core machines or clusters with multiple computers are available, one can use our block-based design to achieve essentially a speedup factor controlled by the number of available cores, assuming each core has access to a memory space that can store one block. In an ideal case with infinitely available cores, the time complexity is dropped to $O(b_s^3)$, which is entirely determined by a single block size.

\section{Validation and application}\label{Sec:Validation}

We performed our tests on a PC with Intel(R) Xeon(R) CPU E5-1630 v3 @ 3.70GHz and 8GB memory. For parallel computing, we used OpenMP (\url{https://computing.llnl.gov/tutorials/openMP/}). We first tested our algorithm on an extremely large multiprotein complex to verify the capability of our algorithm. Then, we analyze the impact of using different block sizes and numbers of threads in terms of the execution time and memory footprint at various grid spacings. Based on the resulting statistics with varying parameter settings, we have confirmed empirically that the memory footprint and execution time indeed behaved as predicted in our analysis above.

All molecular structures used in our validation were downloaded from Protein Data Bank (PDB, \url{https://www.rcsb.org/}). The protein-ligand complexes   used in our application were obtained from PDBbing  (\url{http://www.pdbbind.org.cn/}). All structures were processed with   pdb-to-xyzr   (\href{https://github.com/Pymol-Scripts/Pymol-script-repo/blob/master/modules/MSMS/i64Linux2/pdb_to_xyzr}
{\texttt{https://github.com/Pymol-Scripts/Pymol-script-repo/blob/master/modules/\linebreak MSMS/i64Linux2/pdb\string_to\string_xyzr}}) to assume appropriate van der Waals radii in addition to   atomic coordinates.

\begin{figure}
	\centering
	\includegraphics[width=0.9\textwidth]{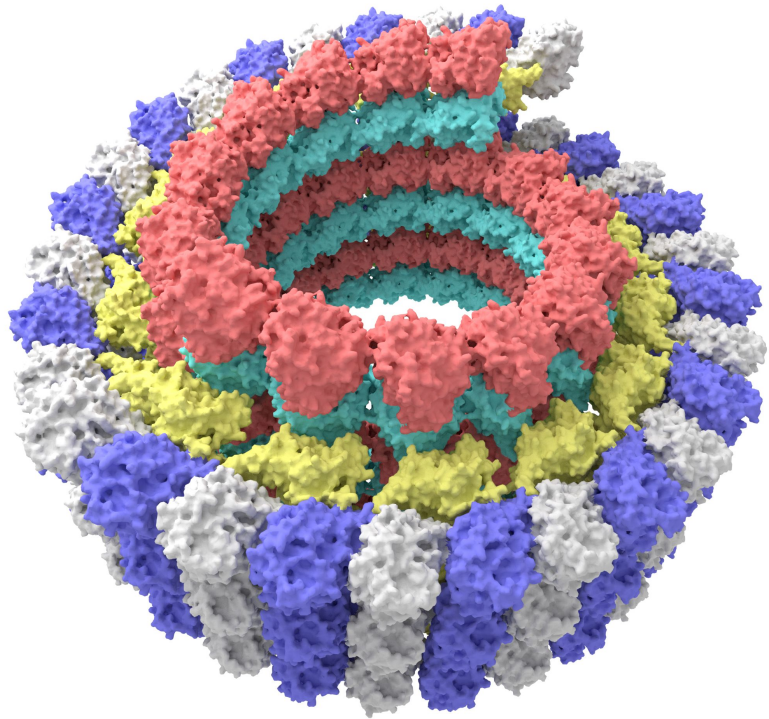}
	\caption{\textbf{Illustration of the SES generation of multiprotein complex.} Here we show the  assembly  of  3j2u proteins with different chains being marked by different colors. Inner ring: microtubule; intermediate ring:  kinesin-13 head domain; and outer ring: curved tubulin protofilament. }
	\label{fig:3u2j_chain}
\end{figure}

\begin{figure}
	\centering
	\includegraphics[width=0.7\textwidth]{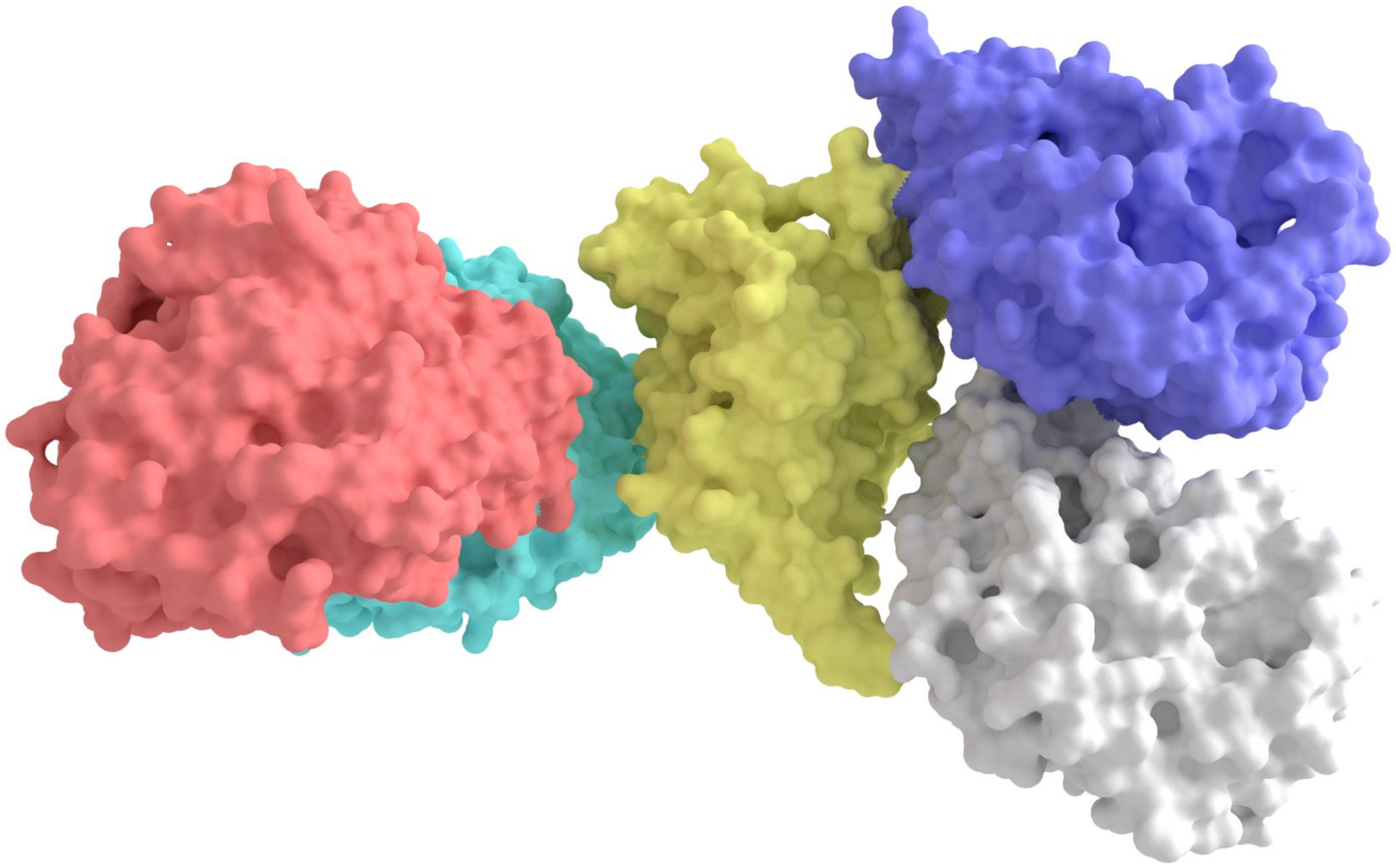}
	\caption{\textbf{Illustration of  the building block protein 3j2u.} It shows Drosophila melanogaster kinesin-13 head domain (yellow) binding to tubulin protofilament (silver and blue) and  microtubule (red and green).   }
	\label{fig:3u2j_small}
\end{figure}
\subsection{Validation on multiprotein complex}

In our tests, the algorithm was able to produce the SES successfully for multiprotein complexes with an arbitrary list of atoms at a very high resolution. For instance, Figure~\ref{fig:3u2j_chain} shows a multiprotein complex consisting of tubulin,  Drosophila melanogaster kinesin-13 KLP10A and  microtubule \cite{asenjo2013structural} {constructed by} protein 3j2u {with 15575 atoms} shown in Figure~\ref{fig:3u2j_small} as the building block. There are 42 such blocks plotted in this multiprotein complex. This typical protein assembly is crucial for investigating the recognition and deformation of tubulins in a microtubule. Due to its excessively large size, such a complex is always an obstacle to handle in theoretical modeling. However, with our software, by only assigning 8 threads to perform the block-based tasks in parallel, the combined memory footprint is controlled to the reasonable amount of 2GB. The whole procedure took about 10 minutes to generate the SES output, including the grid point classification and the intersection information. We were also able to mark different chains in the large protein assembly in the process, with {auxiliary} information provided by our algorithm, which is the nearby atoms of intersections points. This demonstrates the versatility of our algorithm when used in downstream applications, such as solving the Poisson-Boltzmann equation for electrostatic analysis  \cite{li2016multiscale}.

\begin{figure}
	\centering
	\includegraphics[width=0.99\textwidth]{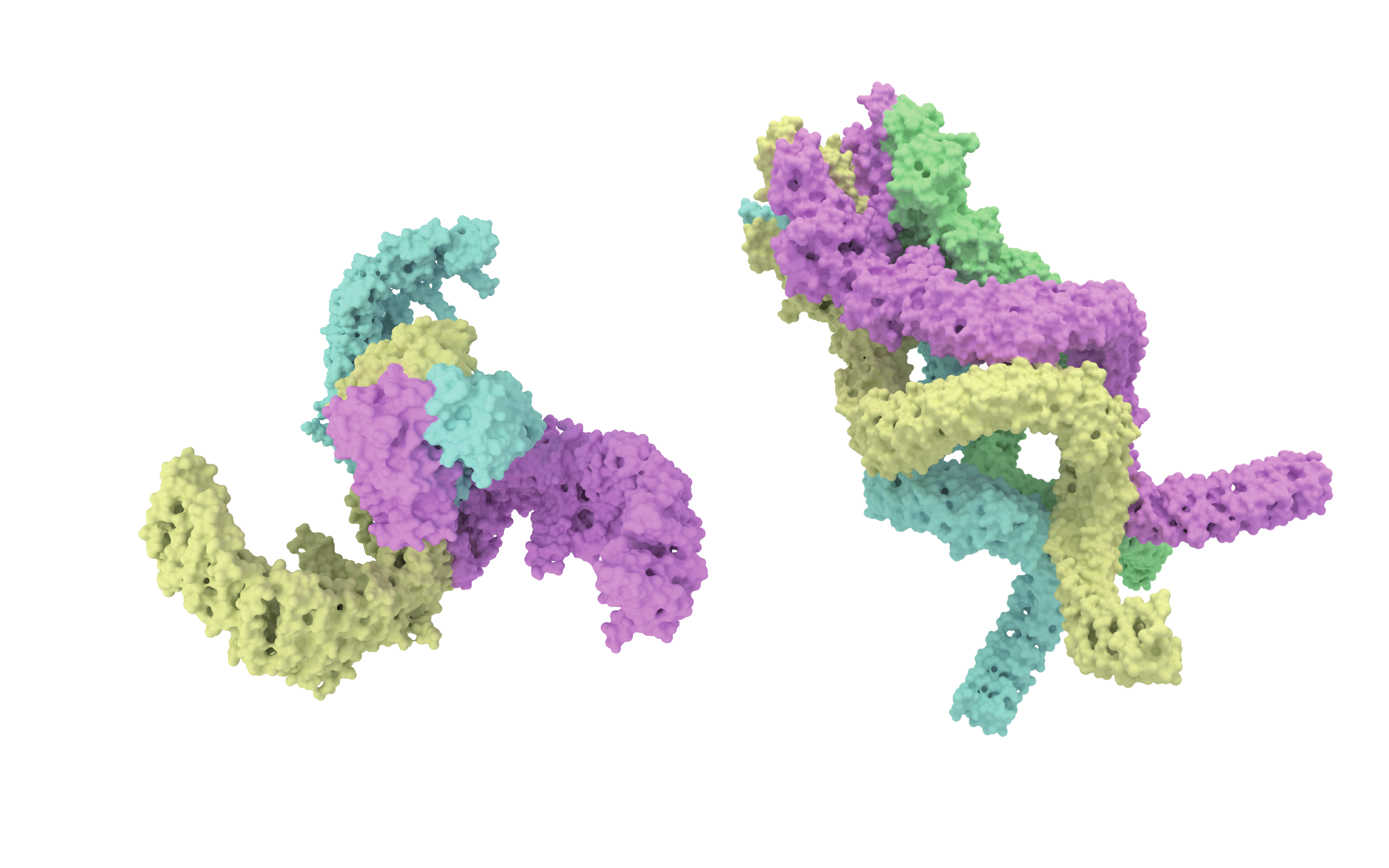}
	\caption{Two models, proteins 5z10 (left) and 5vkq (right), on which tests and statistics of the present algorithm are performed. }
	\label{Fig:two}
\end{figure}

\subsection{Single-thread analysis}

 Protein 5z10, {shown in Fig. \ref{Fig:two} left}, reported by \cite{zhao2018structure} is tested as an example for single-thread performance. This protein is a typical mechanosensitive  ion channel constructed by three identical blade-like subunits. It is found that by probing the state of surrounding membrane, the channel opens with the distortion of these three blades.
	
\begin{figure}
	\centering
	\includegraphics[width=0.8\textwidth]{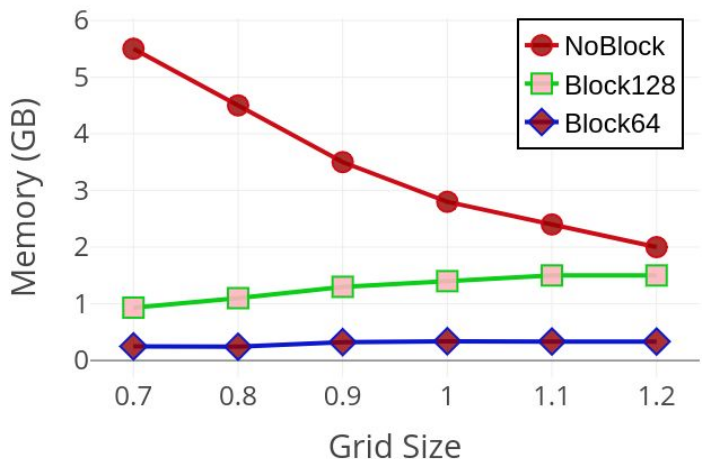}
	\caption{\textbf{Memory footprint comparison at various grid spacings (\AA) for protein 5z10 (single-thread).} Curves with circular, square and diamond nodes are corresponding to no block assigned, block size 128 assignment, and block size 64 assignment, respectively.}
	\label{Fig:singleThreadMem}
\end{figure}

We provide some memory footprint statistics in Figure~\ref{Fig:singleThreadMem} with only a single thread launched. As the plot indicates, if we do not incorporate the block design and use the original method with the entire grid residing in the memory at all time, the memory footprint increases for fine grids, shown by the curve with circular nodes. If we use blocks with size $b_s=128$, the memory footprint drops significantly, simply because only a single block needs to reside in the memory for the single thread, in addition to other auxiliary information, as shown by the curve with square nodes. For the block size $b_s=64$, the curve with diamond nodes shows that the memory footprint is further reduced. By assigning blocks, the memory footprint is dominated and restricted by the information stored within a single block, which is controlled only by the block size and is independent of the grid sizes. Therefore, memory footprint of our approach is well-controlled.

\begin{figure}
	\centering
	\includegraphics[width=0.8\textwidth]{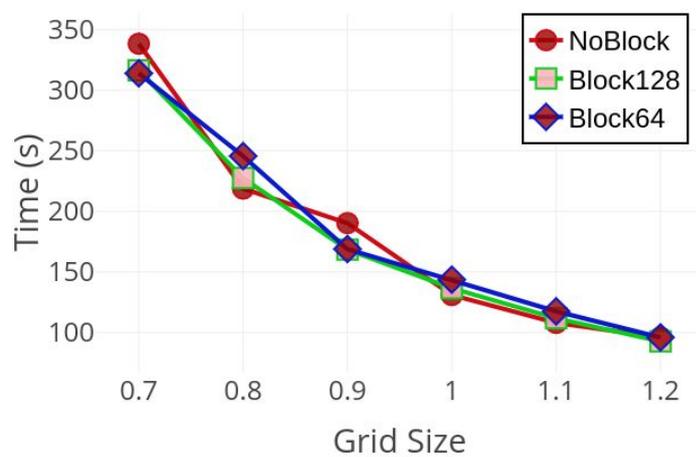}
	\caption{\textbf{Execution time comparison at various grid spacings (\AA) for protein 5z10 (single-thread).} Curves with circular, square and diamond nodes are corresponding to no block assigned, block size 128 assignment, and block size 64 assignment, respectively.
	}
	\label{Fig:singleThreadTime}
\end{figure}

The statistics of execution time is provided in Figure~\ref{Fig:singleThreadTime} for molecule 5z10 with only a single thread. The execution time for different block sizes did not vary significantly. This behavior is expected, since we did not change the total amount of the calculation, for the single-thread version, only the memory footprint is reduced. Note that change was made to the grid point classification or grid edge intersection detection parts. Stated differently, there is no change in analytical nature of the original ESES algorithm.

\subsection{Multi-thread analysis}

Protein 5vkq, {shown in Fig. \ref{Fig:two} right}, reported by \cite{jin2017electron} is used as an example to test multi-thread performance. This protein is a typical mechanosensitive ion channel in bacteria. Its long and spring shaped domains are tethered with microtubules, which will open when it senses the motion of the cytoskeleton environment.

\begin{figure}
	\centering
	\includegraphics[width=0.8\textwidth]{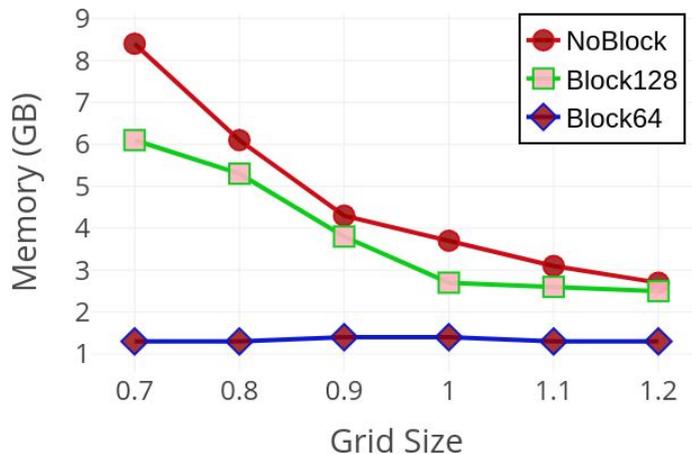}
	\caption{\textbf{Memory footprint comparison at various grid spacings (\AA)  for protein 5vkq (8-threads).} Curves with circular, square and diamond nodes are corresponding to no block assigned, block size 128 assignment, and block size 64 assignment, respectively.}
	\label{Fig:8ThreadMem}
\end{figure}

If we initiate $N$ threads in parallel, we expect that the memory footprint is roughly $N$ times the numbers when the number of $N$ is large. In practice, we observed that it is actually smaller due to that some of the overhead is shared by the threads, especially when $N$ is small. In Figure~\ref{Fig:8ThreadMem}, we present the statistics of memory footprint for protein 5vkq with 8 threads launched in parallel. The curve with circular nodes serves as a baseline when we stick to the original ESES algorithm, which shows an excessive memory requirement. When launching 8 threads, obviously the memory footprint shifted higher compared to launching a single thread simply because we need to load the execution context for all 8 blocks. Nevertheless the memory footprint is still significantly reduced compared to the baseline, unless the number of threads matches the number of blocks. In addition, we still control the memory footprint by the number of threads launched. The curve with square nodes shows the memory usage with block size $b_s=128$, and the curve with diamond nodes shows the case with block size $b_s=64$.

\begin{figure}
	\centering
	\includegraphics[width=0.8\textwidth]{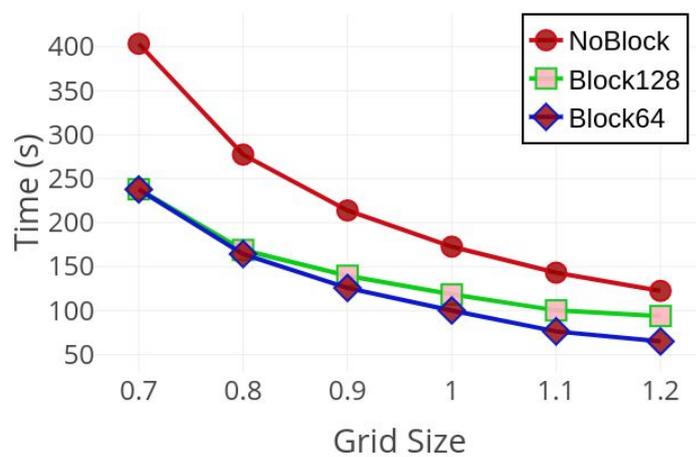}
	\caption{\textbf{Execution time comparison at various grid spacings (\AA)  for protein 5vkq (8-threads).} Curves with circular, square and diamond nodes are corresponding to no block assigned, block size 128 assignment, and block size 64 assignment, respectively.}
	\label{Fig:8ThreadTime}
\end{figure}

The execution time statistics for the same 8-thread experiment for protein 5vkq are shown in Figure~\ref{Fig:8ThreadTime}. The curve with circular nodes gives us a baseline when we stick to the original ESES algorithm. In this example, the execution time was reduced significantly simply by launching several threads at the same time. It is not a perfect 8-times improvement, as predicted by Amdahl's Law \cite{mccool2012structured}, because there are always critical sections that need serial execution such as file I/O. We also found that a smaller block size (curve with diamond nodes) also brings some additional improvements as in the single-thread mode. It is most likely due to the same reason that the memory allocation is easier for smaller blocks. Taking into account both the spatial and temporal statistics, we observed that by reducing the block size, we can significantly reduce the memory footprint without any negative impact on the time performance and algorithm accuracy.

Finally, we apply the present approach to a large set of protein-ligand complexes. We consider the  PDBbing v2016 core set of 290 protein-ligand complexes. Our  results in terms of grid dimension, block dimension, surface area, and surface enclosed volume are given for each protein in {Appendix~\ref{apdx:list}}. These results can be used by independent researchers to validate their own surface generations. The computational  parameters are set to probe size 1.4\AA, grid spacing 0.4\AA, grid extension 0.8\AA~ and block size 64. Note that the proposed method has no effect on the ESES generation quality. {The proposed method can thus be used as an efficient replacement to ESES, and be applied to any solvent excluded surface based molecular modeling and analysis.}

\begin{figure}
	\centering
	\includegraphics[width=0.8\textwidth]{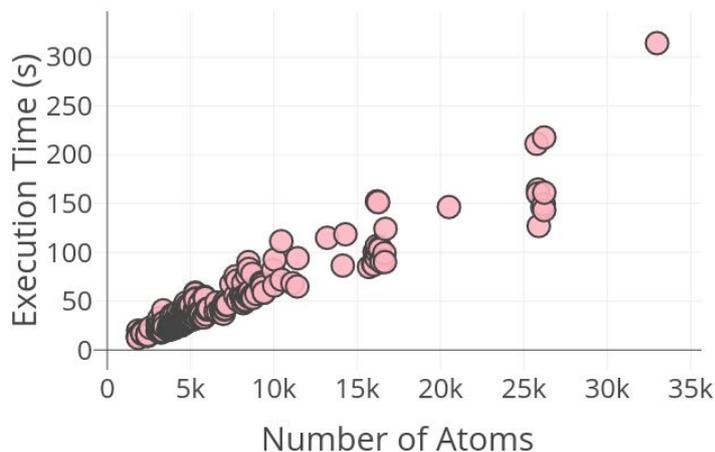}
	\caption{{ \textbf{Execution time analysis.} The scatter plot of execution time vs the number of atoms is given for  290 proteins when   8 threads are used.}}
	\label{Fig:timePlot}
\end{figure}

\section{Conclusion}\label{Sec:Conclusion}
We present a divide-and-conquer approach to solve the memory explosion issue when dealing with large macromolecules at high grid resolutions. The approach is based on the localized nature of the computations involved in Eulerian solute-excluded surfaces (ESESs) \cite{JCC:JCC24682}. In the present approach, we partition the entire computational domain into subdomains (blocks) that can fit into a given size of memory space. The memory requirement is determined  by the data in the block(s) used in the current calculation. In this manner, we can control the upper bound of the memory footprint, and allow the user to run our software on a typical commodity personal computer (PC). Taking the advantage of the locality, we  also incorporate the power of parallel computing to further enhance the performance. With such a practical implementation, we can significantly  extend the applicability of  the earlier ESES algorithm by lifting its constraints of memory requirements and running on a single CPU. The present improvement does not change the analytical nature of the original ESES algorithm. The proposed method is validated on the ESES generation of an excessively large protein complex and a couple of large proteins. Application is considered to 290 protein-ligand complexes.

There is still a room for further improvements. In the potential analytical patch construction, especially for the saddle and concave patches, we simply consider all possible pairs of atoms which are at a distance below a threshold determined by the van der Waals radii. Apparently, there is a redundancy in such an approach, since some patches are buried inside the molecular surface and may be pre-culled to save computation. As future work, we wish to explore fast calculations that can eliminate such patches before classifying grid points. Another direction to explore is to consider GPU computing, since a similar parallelized design can be applied when mapping them to GPU threads and blocks instead of CPU cores. A central issue in carrying out a GPU implementation is how the analytical SES patch construction and the associated high order polynomial root finding at grid edge and surface intersection can be efficiently adapted to the less powerful ALU units on GPUs. Further simplifications may be desirable to harness the power of GPU for this problem.

 \section*{Acknowledgments}
{This work was supported in part by NSF Grants DMS-1721024, DMS-1761320, and CCF-1655422, and NIH grant 1R01GM126189-01A1.}


\begin{appendices}
	
\section{Surface generation test}

\begin{longtable}{ |p{1.7cm}||p{1.5cm}|p{2cm}|p{1.8cm}|p{1.2cm}|p{1.8cm}|p{1.2cm}|  }
	\hline
	Protein ID & \#Atoms & Grid Dim & Block Dim & Area(\AA$^2$) & Volume(\AA$^3$) & time($s$) \\
	\hline
	3dxg & 1856 & 105x101x117 & 2x2x2 & 5777.88 & 16569.2 & 13.0162 \\
	3d6q & 1856 & 105x102x117 & 2x2x2 & 5793.47 & 16601.3 & 19.8521 \\
	1w4o & 1856 & 105x99x116 & 2x2x2 & 5776.42 & 16575.4 & 14.2584 \\
	1o0h & 1856 & 104x108x117 & 2x2x2 & 5753.33 & 16456.4 & 14.2019 \\
	1u1b & 1856 & 103x129x94 & 2x3x2 & 5790.07 & 16722.8 & 12.3264 \\
	4lzs & 2121 & 95x132x127 & 2x3x2 & 6585.37 & 18644.5 & 19.2717 \\
	3u5j & 2121 & 95x135x127 & 2x3x2 & 6686.26 & 18677.5 & 17.7138 \\
	4wiv & 2121 & 138x85x125 & 3x2x2 & 6749.24 & 18694.1 & 16.3995 \\
	4ogj & 2121 & 129x110x110 & 3x2x2 & 6710.66 & 18851.7 & 16.8964 \\
	3p5o & 2121 & 96x135x125 & 2x3x2 & 6718.7 & 18764.6 & 17.1206 \\
	3lka & 2408 & 110x112x101 & 2x2x2 & 7040.19 & 21685.7 & 16.9318 \\
	3ehy & 2408 & 110x112x100 & 2x2x2 & 7039.49 & 21574.7 & 15.2638 \\
	3nx7 & 2408 & 112x112x101 & 2x2x2 & 6960.15 & 21681.9 & 14.4231 \\
	3tsk & 2425 & 109x100x119 & 2x2x2 & 7057.27 & 22082.1 & 14.6475 \\
	4gr0 & 2425 & 109x101x115 & 2x2x2 & 6962.62 & 21868.1 & 15.4103 \\
	3nq9 & 2584 & 120x120x112 & 2x2x2 & 7368.2 & 23464 & 22.7301 \\
	5aba & 3025 & 111x126x136 & 2x2x3 & 8747.75 & 27568.7 & 25.6786 \\
	4agq & 3025 & 113x126x138 & 2x2x3 & 8885.8 & 27571.7 & 19.5389 \\
	5a7b & 3047 & 111x130x136 & 2x3x3 & 8943.39 & 27637.8 & 27.9725 \\
	4agp & 3047 & 114x126x138 & 2x2x3 & 8877.19 & 27718.7 & 27.5211 \\
	4agn & 3069 & 112x126x137 & 2x2x3 & 8966.61 & 27884.4 & 29.8699 \\
	2qnq & 3128 & 104x139x143 & 2x3x3 & 8577.43 & 27855.4 & 20.697 \\
	3cyx & 3128 & 106x134x142 & 2x3x3 & 8566.31 & 27591.5 & 21.0516 \\
	1eby & 3134 & 107x136x147 & 2x3x3 & 8604.66 & 28389.9 & 21.5876 \\
	3o9i & 3134 & 152x103x111 & 3x2x2 & 8526.46 & 27838 & 19.3083 \\
	1a30 & 3138 & 107x136x140 & 2x3x3 & 8459.42 & 27734.2 & 32.6073 \\
	4abg & 3204 & 109x118x135 & 2x2x3 & 8104.51 & 29113.2 & 27.3888 \\
	1uto & 3220 & 123x99x132 & 2x2x3 & 8037.53 & 29010 & 17.8095 \\
	3gy4 & 3220 & 123x107x133 & 2x2x3 & 7993.9 & 29061.3 & 20.5542 \\
	1k1i & 3220 & 124x108x131 & 2x2x3 & 8127.1 & 29250 & 20.144 \\
	1o3f & 3220 & 121x132x111 & 2x3x2 & 8059.72 & 29193.8 & 19.3656 \\
	3kr8 & 3248 & 140x132x106 & 3x3x2 & 9356.53 & 29196.6 & 20.4359 \\
	2yki & 3259 & 122x124x126 & 2x2x2 & 9171.29 & 29287.1 & 19.6484 \\
	4kzq & 3278 & 137x130x103 & 3x3x2 & 9356.27 & 29733.6 & 19.8373 \\
	4kzu & 3278 & 139x131x108 & 3x3x2 & 9308.97 & 29802.4 & 20.4075 \\
	4j21 & 3292 & 149x125x107 & 3x2x2 & 9533.27 & 30100.4 & 20.4142 \\
	4j3l & 3292 & 125x150x106 & 2x3x2 & 9398.54 & 29849.4 & 19.9642 \\
	1yc1 & 3313 & 119x110x129 & 2x2x3 & 9072.47 & 29432.3 & 20.9225 \\
	3ozt & 3357 & 200x148x114 & 4x3x2 & 10015 & 30240.5 & 40.6189 \\
	3ozs & 3357 & 196x150x114 & 4x3x2 & 9996.25 & 30152.6 & 26.3215 \\
	3oe5 & 3357 & 193x149x113 & 4x3x2 & 9925.33 & 30107 & 27.2306 \\
	3oe4 & 3357 & 193x149x114 & 4x3x2 & 9980.64 & 30163 & 27.2095 \\
	3nw9 & 3365 & 107x115x144 & 2x2x3 & 8131.95 & 30170.6 & 19.38 \\
	3b27 & 3388 & 128x126x159 & 3x2x3 & 9731.4 & 30503.3 & 24.2988 \\
	2fxs & 3409 & 151x128x132 & 3x3x3 & 9203.98 & 30434.6 & 23.9637 \\
	2yge & 3420 & 151x130x134 & 3x3x3 & 9285.46 & 30724.4 & 27.1157 \\
	2iwx & 3426 & 155x129x133 & 3x3x3 & 9345.14 & 30760.3 & 27.0977 \\
	2vw5 & 3426 & 157x129x134 & 3x3x3 & 9341.02 & 30901.6 & 28.3237 \\
	3rlr & 3449 & 121x121x156 & 2x2x3 & 9914.87 & 30865.4 & 23.8025 \\
	1lpg & 3665 & 125x136x114 & 2x3x2 & 9706.24 & 32892.4 & 20.5021 \\
	4crc & 3711 & 127x120x141 & 2x2x3 & 9907.48 & 33608.9 & 22.8678 \\
	4x6p & 3715 & 140x116x126 & 3x2x2 & 10131.4 & 33827.4 & 20.8315 \\
	4cra & 3724 & 145x128x115 & 3x3x2 & 9972.72 & 33624 & 22.1204 \\
	4ty7 & 3727 & 133x138x144 & 3x3x3 & 10136.1 & 33908.5 & 24.43 \\
	3kgp & 3737 & 144x112x123 & 3x2x2 & 9519.34 & 33612.6 & 23.194 \\
	1o5b & 3810 & 146x127x120 & 3x2x2 & 9445.63 & 34582.3 & 22.1093 \\
	1c5z & 3811 & 146x129x120 & 3x3x2 & 9531.12 & 34644.1 & 33.7694 \\
	1sqa & 3818 & 128x157x125 & 3x3x2 & 10141.1 & 35162.1 & 23.7716 \\
	1owh & 3820 & 127x156x125 & 2x3x2 & 10149.2 & 35375.6 & 23.6066 \\
	4qd6 & 3849 & 134x147x211 & 3x3x4 & 12029.2 & 35089.4 & 33.5549 \\
	3qgy & 3896 & 133x135x159 & 3x3x3 & 10853.7 & 35113.4 & 28.1767 \\
	4de2 & 3900 & 138x109x147 & 3x2x3 & 9295.39 & 34595.6 & 35.9985 \\
	4de3 & 3900 & 138x111x144 & 3x2x3 & 9241.18 & 34660.7 & 23.651 \\
	1z95 & 3917 & 141x120x156 & 3x2x3 & 10169.7 & 35038.8 & 27.3216 \\
	4de1 & 3923 & 139x108x152 & 3x2x3 & 9427.62 & 34862.4 & 23.8837 \\
	3g2z & 3925 & 140x109x152 & 3x2x3 & 9285.38 & 34891.5 & 22.2739 \\
	3g31 & 3925 & 140x110x147 & 3x2x3 & 9326.88 & 34989 & 26.6228 \\
	4rfm & 3942 & 138x128x154 & 3x3x3 & 10899.5 & 35830 & 25.4256 \\
	3kwa & 4032 & 127x125x150 & 2x2x3 & 10003.9 & 36326.8 & 35.8207 \\
	4jsz & 4048 & 127x124x149 & 2x2x3 & 10121.1 & 36590 & 22.4405 \\
	3ryj & 4054 & 129x123x148 & 3x2x3 & 10333.2 & 36435.3 & 23.7567 \\
	3b68 & 4068 & 149x120x158 & 3x2x3 & 10827.6 & 36233.7 & 29.5425 \\
	3b5r & 4068 & 149x120x159 & 3x2x3 & 10866.6 & 36325.1 & 28.6952 \\
	3b65 & 4068 & 149x120x158 & 3x2x3 & 10801 & 36251.7 & 28.1824 \\
	2weg & 4071 & 128x121x147 & 3x2x3 & 10146 & 36513.5 & 24.3336 \\
	3dd0 & 4071 & 130x123x149 & 3x2x3 & 10322.6 & 36951.8 & 23.7389 \\
	3gbb & 4086 & 123x149x157 & 2x3x3 & 11431.4 & 36652.7 & 27.8599 \\
	3g0w & 4099 & 153x124x150 & 3x2x3 & 10647.3 & 36712.6 & 27.9932 \\
	3fv2 & 4101 & 145x134x121 & 3x3x2 & 11330.7 & 36688 & 24.859 \\
	3fv1 & 4101 & 145x140x122 & 3x3x2 & 11270.2 & 36526.1 & 24.8372 \\
	3fur & 4161 & 133x110x172 & 3x2x3 & 11796.3 & 37539 & 25.4867 \\
	3myg & 4169 & 156x145x126 & 3x3x2 & 11063 & 37469.2 & 27.555 \\
	4m0y & 4177 & 131x152x148 & 3x3x3 & 11990.4 & 38728.9 & 29.0143 \\
	3u9q & 4179 & 124x131x154 & 2x3x3 & 11740 & 37330.8 & 33.5672 \\
	3jvs & 4186 & 139x118x183 & 3x2x3 & 11445 & 37468.2 & 27.3559 \\
	4m0z & 4199 & 128x148x149 & 3x3x3 & 11655.2 & 38361.3 & 30.2577 \\
	3ao4 & 4210 & 137x133x144 & 3x3x3 & 10992 & 38287.1 & 26.372 \\
	3jvr & 4218 & 138x116x181 & 3x2x3 & 11495.1 & 37840.2 & 24.8808 \\
	3b1m & 4229 & 116x129x169 & 2x3x3 & 11722.6 & 37809.4 & 27.7153 \\
	3mss & 4263 & 131x118x172 & 3x2x3 & 12047.8 & 38810.3 & 38.2454 \\
	2c3i & 4274 & 154x144x123 & 3x3x2 & 11301.2 & 38876.8 & 26.4624 \\
	2yfe & 4281 & 139x156x153 & 3x3x3 & 11923.7 & 38510.9 & 31.6776 \\
	3pyy & 4289 & 111x128x166 & 2x3x3 & 11972.6 & 38784.4 & 25.2828 \\
	1nvq & 4291 & 139x124x181 & 3x2x3 & 11909.9 & 38394.9 & 28.3015 \\
	2xbv & 4291 & 149x135x131 & 3x3x3 & 11794.8 & 39406.4 & 27.6058 \\
	4twp & 4292 & 165x147x113 & 3x3x2 & 11511.7 & 38978.3 & 27.6603 \\
	3bgz & 4318 & 158x141x121 & 3x3x2 & 11265.8 & 39021.6 & 28.2811 \\
	2wtv & 4337 & 118x161x160 & 2x3x3 & 11713 & 38572.4 & 27.7694 \\
	2v7a & 4340 & 129x136x166 & 3x3x3 & 12063.1 & 39932 & 28.3806 \\
	1mq6 & 4372 & 167x139x112 & 3x3x2 & 11625.3 & 39651.5 & 26.7912 \\
	3uo4 & 4392 & 157x142x139 & 3x3x3 & 11569.5 & 39341.1 & 32.1715 \\
	3up2 & 4392 & 158x145x139 & 3x3x3 & 11612.7 & 39524.3 & 31.3933 \\
	5dwr & 4397 & 135x151x134 & 3x3x3 & 11467.8 & 40084.1 & 27.2752 \\
	3jya & 4408 & 134x155x133 & 3x3x3 & 11571 & 40232.4 & 27.4488 \\
	1z6e & 4413 & 164x142x109 & 3x3x2 & 11656.5 & 39643.9 & 25.5687 \\
	2brb & 4425 & 141x133x182 & 3x3x3 & 12414 & 40414.9 & 31.3865 \\
	2br1 & 4425 & 138x135x183 & 3x3x3 & 12257.1 & 39755.9 & 32.0098 \\
	3utu & 4434 & 128x130x139 & 3x3x3 & 11221.5 & 39829.2 & 24.9517 \\
	2y5h & 4440 & 140x168x116 & 3x3x2 & 12126 & 40601.4 & 26.7111 \\
	1bcu & 4446 & 128x129x140 & 3x3x3 & 11130.7 & 40337.1 & 37.4077 \\
	4k18 & 4453 & 161x141x132 & 3x3x3 & 11652.3 & 40461.1 & 28.9929 \\
	1oyt & 4479 & 129x132x140 & 3x3x3 & 11485.2 & 40270.1 & 26.678 \\
	2zda & 4513 & 131x131x143 & 3x3x3 & 11344.7 & 40606 & 27.9352 \\
	4k77 & 4549 & 146x142x150 & 3x3x3 & 12696.8 & 41470.2 & 30.1223 \\
	4cig & 4564 & 144x125x152 & 3x2x3 & 12304.1 & 41078.4 & 40.9304 \\
	3k5v & 4588 & 143x184x124 & 3x3x2 & 13284.3 & 41730.7 & 29.958 \\
	3bv9 & 4621 & 138x138x136 & 3x3x3 & 11636.2 & 41530.9 & 27.8745 \\
	3zt2 & 4642 & 142x131x155 & 3x3x3 & 12440.7 & 41627.9 & 32.2067 \\
	3zsx & 4642 & 144x130x153 & 3x3x3 & 12556.8 & 41830 & 45.4165 \\
	2zy1 & 4646 & 149x144x159 & 3x3x3 & 12283.4 & 41678.4 & 31.8607 \\
	3uri & 4650 & 147x152x134 & 3x3x3 & 11386.5 & 42463.2 & 29.2076 \\
	2fvd & 4653 & 140x114x170 & 3x2x3 & 12820 & 41544.9 & 30.1102 \\
	2v00 & 4669 & 139x127x166 & 3x2x3 & 11462.2 & 42410.6 & 42.6454 \\
	3wz8 & 4669 & 140x127x167 & 3x2x3 & 11460.9 & 42536.5 & 27.7162 \\
	3pww & 4669 & 138x126x166 & 3x2x3 & 11500.2 & 42377.7 & 30.2387 \\
	3prs & 4669 & 139x125x166 & 3x2x3 & 11499.9 & 42426.3 & 28.767 \\
	3zso & 4670 & 143x132x153 & 3x3x3 & 12690 & 41991.9 & 31.6446 \\
	4ea2 & 4670 & 148x144x161 & 3x3x3 & 12343.4 & 42173 & 32.4862 \\
	2zcr & 4670 & 149x147x160 & 3x3x3 & 12337 & 41938.9 & 33.0923 \\
	4hge & 4674 & 161x156x143 & 3x3x3 & 13043.1 & 42660 & 33.0706 \\
	4ivd & 4698 & 214x143x158 & 4x3x3 & 13788.6 & 42619.9 & 40.4818 \\
	3fcq & 4700 & 173x131x132 & 3x3x3 & 10956.1 & 42376.4 & 34.77 \\
	1z9g & 4700 & 175x132x133 & 3x3x3 & 11019.2 & 42836.4 & 29.4944 \\
	1qf1 & 4700 & 173x132x132 & 3x3x3 & 10908.5 & 42623.9 & 28.7241 \\
	4ivb & 4713 & 215x143x158 & 4x3x3 & 13772.8 & 42947.1 & 41.5465 \\
	4ivc & 4713 & 213x142x158 & 4x3x3 & 13712.4 & 42361 & 40.4934 \\
	3acw & 4714 & 150x146x161 & 3x3x3 & 12637.4 & 42325.7 & 48.4493 \\
	2zcq & 4714 & 145x143x157 & 3x3x3 & 12308.2 & 42281.4 & 31.2949 \\
	4f09 & 4739 & 143x159x169 & 3x3x3 & 13163 & 42812.4 & 35.2086 \\
	4gfm & 4762 & 145x148x129 & 3x3x3 & 12925.4 & 43402.4 & 28.8006 \\
	1pxn & 4788 & 142x114x170 & 3x2x3 & 13001.5 & 42799.9 & 30.9308 \\
	2hb1 & 4811 & 170x145x117 & 3x3x2 & 11635.3 & 43180.8 & 45.4613 \\
	1bzc & 4811 & 173x142x115 & 3x3x2 & 11505.4 & 43489.1 & 30.1936 \\
	2qbq & 4811 & 170x145x118 & 3x3x2 & 11678.6 & 43351.1 & 31.5649 \\
	2qbp & 4811 & 169x145x117 & 3x3x2 & 11655.4 & 43424.7 & 31.878 \\
	2xnb & 4819 & 141x111x171 & 3x2x3 & 12885.2 & 43258.1 & 29.5643 \\
	2qbr & 4830 & 177x145x120 & 3x3x2 & 11737.8 & 43724.4 & 30.8117 \\
	3e5a & 4850 & 152x161x145 & 3x3x3 & 13072.8 & 43625.6 & 32.6783 \\
	4e6q & 4869 & 168x138x181 & 3x3x3 & 13820.9 & 43831.2 & 37.587 \\
	4jia & 4878 & 142x183x165 & 3x3x3 & 14222.9 & 44267.3 & 37.1466 \\
	3pxf & 4908 & 157x110x180 & 3x2x3 & 13841.6 & 43820.6 & 43.9436 \\
	3uuo & 4922 & 145x150x157 & 3x3x3 & 12998.8 & 45572.7 & 31.3175 \\
	3ueu & 4966 & 123x130x203 & 2x3x4 & 13255.2 & 44763.8 & 32.4677 \\
	3uew & 5000 & 123x128x213 & 2x3x4 & 13666.4 & 45435.3 & 33.8787 \\
	3twp & 5009 & 164x127x133 & 3x2x3 & 12462.3 & 45286.2 & 45.8088 \\
	3qqs & 5009 & 134x166x147 & 3x3x3 & 12412.7 & 45349.6 & 31.8109 \\
	3uev & 5068 & 125x128x201 & 2x3x4 & 13640.4 & 45616.3 & 31.0619 \\
	3ui7 & 5091 & 142x151x157 & 3x3x3 & 13135.4 & 46772.9 & 32.3145 \\
	3uex & 5096 & 124x129x209 & 2x3x4 & 13975.3 & 46266 & 32.1042 \\
	5c2h & 5191 & 151x147x149 & 3x3x3 & 13072.6 & 47126.9 & 35.0963 \\
	5c1w & 5235 & 147x156x157 & 3x3x3 & 13216.2 & 47644.6 & 36.0221 \\
	5c28 & 5235 & 148x154x154 & 3x3x3 & 13159.3 & 47569.4 & 32.2369 \\
	3ag9 & 5267 & 137x174x143 & 3x3x3 & 13473.3 & 47423.8 & 36.549 \\
	4w9h & 5271 & 174x143x199 & 3x3x4 & 14645.7 & 47401.8 & 41.345 \\
	4msc & 5280 & 147x158x158 & 3x3x3 & 13787.5 & 48439.7 & 34.8642 \\
	4w9c & 5282 & 175x145x200 & 3x3x4 & 14685.8 & 47924.2 & 58.7774 \\
	4w9l & 5282 & 174x145x197 & 3x3x4 & 14812.9 & 47975.6 & 41.5329 \\
	4w9i & 5282 & 177x143x199 & 3x3x4 & 14614.3 & 47854.5 & 40.2372 \\
	4bkt & 5296 & 141x179x201 & 3x3x4 & 14714.9 & 48335 & 57.7965 \\
	4llx & 5299 & 150x159x157 & 3x3x3 & 13540.6 & 48052.3 & 38.9411 \\
	4mrw & 5299 & 150x159x157 & 3x3x3 & 13525.4 & 48193.6 & 50.8482 \\
	4mrz & 5299 & 149x160x156 & 3x3x3 & 13504.9 & 47956.9 & 52.8705 \\
	4msn & 5299 & 150x159x155 & 3x3x3 & 13688.1 & 48625.8 & 51.905 \\
	4dli & 5411 & 185x140x133 & 3x3x3 & 15130.5 & 49108.4 & 36.0334 \\
	4f9w & 5433 & 185x140x129 & 3x3x3 & 14761.8 & 48989.7 & 33.0079 \\
	2zb1 & 5575 & 155x149x155 & 3x3x3 & 15142.3 & 50414.7 & 37.3719 \\
	3gv9 & 5581 & 153x144x145 & 3x3x3 & 12928.2 & 50191.6 & 33.2439 \\
	3gr2 & 5581 & 153x140x144 & 3x3x3 & 12961.9 & 50214 & 35.9617 \\
	4kz6 & 5581 & 151x156x133 & 3x3x3 & 12812.9 & 49828.8 & 48.0046 \\
	4jxs & 5581 & 151x157x134 & 3x3x3 & 13053.9 & 50220.6 & 47.9802 \\
	2r9w & 5581 & 154x141x143 & 3x3x3 & 13174 & 49905.5 & 34.4525 \\
	3e93 & 5584 & 156x156x159 & 3x3x3 & 15401.5 & 50252.8 & 39.2194 \\
	1r5y & 5595 & 146x134x181 & 3x3x3 & 13256.5 & 50372 & 36.9078 \\
	3e92 & 5609 & 156x152x158 & 3x3x3 & 15307 & 50559.2 & 38.9401 \\
	1s38 & 5650 & 146x132x179 & 3x3x3 & 13268 & 50828.2 & 36.4457 \\
	1ydr & 5813 & 136x178x129 & 3x3x3 & 14353.4 & 52608.3 & 34.2593 \\
	1ydt & 5813 & 136x177x129 & 3x3x3 & 14423 & 52672.4 & 33.32 \\
	3rsx & 5833 & 154x178x132 & 3x3x3 & 14770.9 & 54165.6 & 54.6174 \\
	1q8t & 5875 & 141x186x138 & 3x3x3 & 15349.5 & 54420.4 & 53.2045 \\
	1q8u & 5928 & 141x187x134 & 3x3x3 & 15888.7 & 55111.5 & 36.2449 \\
	4gid & 5997 & 160x173x150 & 3x3x3 & 14746.3 & 54907.2 & 41.5304 \\
	2vkm & 6019 & 160x175x149 & 3x3x3 & 14775.8 & 54738.4 & 41.6591 \\
	4djv & 6034 & 166x149x174 & 3x3x3 & 14975.7 & 55246.1 & 42.4239 \\
	3udh & 6111 & 166x182x131 & 3x3x3 & 15607.2 & 55890.3 & 42.4351 \\
	3wtj & 6504 & 197x116x177 & 4x2x3 & 18210.3 & 58882.8 & 39.4775 \\
	2xdl & 6560 & 131x191x125 & 3x4x2 & 17346.2 & 59278 & 48.6823 \\
	2qe4 & 6651 & 172x149x145 & 3x3x3 & 16615.2 & 59904.7 & 39.4594 \\
	2wer & 6846 & 129x223x147 & 3x4x3 & 17262.1 & 61418.2 & 42.2958 \\
	4f3c & 6854 & 200x136x128 & 4x3x3 & 14854 & 62181.4 & 40.3111 \\
	1nc3 & 6896 & 167x108x189 & 3x2x3 & 15249.6 & 61913 & 39.4786 \\
	1nc1 & 6905 & 168x106x190 & 3x2x3 & 15224.4 & 62500 & 37.7971 \\
	1y6r & 6905 & 169x107x191 & 3x2x4 & 15335.1 & 62804.4 & 37.6239 \\
	4f2w & 6979 & 109x145x201 & 2x3x4 & 15692.2 & 63726.8 & 37.0809 \\
	2cet & 7025 & 156x147x148 & 3x3x3 & 14613.8 & 63522.8 & 41.3852 \\
	4jfs & 7054 & 157x181x185 & 3x3x3 & 16911.8 & 63804.5 & 48.7066 \\
	4j28 & 7054 & 157x184x184 & 3x3x3 & 17110.9 & 64082 & 46.2038 \\
	2xii & 7054 & 160x148x195 & 3x3x4 & 16914 & 64003.5 & 44.4223 \\
	2j7h & 7066 & 155x149x149 & 3x3x3 & 14854.5 & 64039.6 & 42.2341 \\
	4pcs & 7076 & 188x174x177 & 3x3x3 & 17301.3 & 64169.7 & 49.6077 \\
	2cbv & 7141 & 162x161x150 & 3x3x3 & 14978.2 & 64682.5 & 46.0519 \\
	2j78 & 7142 & 163x153x156 & 3x3x3 & 15058.2 & 64684 & 47.1593 \\
	2pog & 7210 & 161x174x166 & 3x3x3 & 18032.6 & 64766.2 & 46.4577 \\
	4cr9 & 7448 & 213x142x171 & 4x3x3 & 19486.5 & 67552.5 & 67.7259 \\
	2p4y & 7688 & 192x177x185 & 4x3x3 & 20497.6 & 69126.1 & 55.2999 \\
	4mgd & 7695 & 167x173x185 & 3x3x3 & 19584.2 & 69673.7 & 75.4981 \\
	1vso & 7794 & 173x213x149 & 3x4x3 & 20774.4 & 69624.7 & 71.1788 \\
	2p15 & 7849 & 172x166x189 & 3x3x3 & 19652.8 & 71004.1 & 50.9455 \\
	1qkt & 8026 & 184x199x165 & 3x4x3 & 21676 & 72259.2 & 56.2978 \\
	4mme & 8118 & 157x182x161 & 3x3x3 & 17939 & 73566.5 & 52.593 \\
	1p1q & 8134 & 188x174x146 & 3x3x3 & 20746 & 72655.6 & 52.0299 \\
	1p1n & 8134 & 191x188x149 & 4x3x3 & 21060.8 & 72683.9 & 55.3808 \\
	4dld & 8158 & 172x162x182 & 3x3x3 & 21683.9 & 73727 & 50.1094 \\
	4u4s & 8162 & 194x214x155 & 4x4x3 & 21407.1 & 73234.4 & 68.8114 \\
	1syi & 8170 & 159x151x182 & 3x3x3 & 21317.4 & 73091.7 & 47.9248 \\
	2al5 & 8186 & 148x199x191 & 3x4x4 & 21975.7 & 73709.8 & 54.9643 \\
	1h23 & 8286 & 174x168x164 & 3x3x3 & 18128.9 & 75779.4 & 49.7802 \\
	1h22 & 8296 & 174x168x159 & 3x3x3 & 17949.9 & 76064.8 & 50.8137 \\
	1gpk & 8301 & 170x171x166 & 3x3x3 & 18075.1 & 75389.9 & 52.4765 \\
	1gpn & 8303 & 171x171x161 & 3x3x3 & 18205.4 & 75497 & 51.2762 \\
	3coy & 8382 & 171x162x236 & 3x3x4 & 21615.4 & 75471.6 & 56.5176 \\
	3ivg & 8393 & 171x164x225 & 3x3x4 & 21473.8 & 75420.7 & 77.236 \\
	3coz & 8472 & 171x163x236 & 3x3x4 & 21908.6 & 76264.1 & 55.5484 \\
	3aru & 8474 & 258x145x182 & 5x3x3 & 19817.1 & 77814.7 & 90.0987 \\
	4ddh & 8495 & 168x163x226 & 3x3x4 & 21761.8 & 76872.6 & 83.0679 \\
	4ddk & 8519 & 169x164x236 & 3x3x4 & 21986.4 & 76989.4 & 54.1794 \\
	3arp & 8549 & 249x158x177 & 4x3x3 & 20026.1 & 78534 & 58.3026 \\
	3arv & 8563 & 249x158x178 & 4x3x3 & 20188.7 & 78669.3 & 56.6699 \\
	3ary & 8563 & 249x159x176 & 4x3x3 & 20110.6 & 78665.2 & 58.0375 \\
	3arq & 8563 & 249x158x177 & 4x3x3 & 20159.2 & 78685.2 & 59.7228 \\
	4eo8 & 8692 & 161x174x194 & 3x3x4 & 21052 & 78285.7 & 55.5408 \\
	4ih7 & 8701 & 164x173x198 & 3x3x4 & 20699.9 & 79327.2 & 53.9337 \\
	4ih5 & 8703 & 168x173x196 & 3x3x4 & 20942.1 & 79392 & 77.9905 \\
	3cj4 & 8720 & 174x168x193 & 3x3x4 & 21129.3 & 78945.2 & 56.0903 \\
	3gnw & 8738 & 162x166x201 & 3x3x4 & 20977.2 & 78570.2 & 53.2703 \\
	4eor & 8949 & 165x203x190 & 3x4x3 & 21622.7 & 82242.2 & 57.4209 \\
	4e5w & 9243 & 234x238x156 & 4x4x3 & 25469.5 & 84609.8 & 69.0594 \\
	2wca & 9295 & 232x207x176 & 4x4x3 & 21978.8 & 84461.4 & 67.2198 \\
	2w4x & 9306 & 230x207x177 & 4x4x3 & 21814.6 & 83862 & 64.4236 \\
	5tmn & 9400 & 213x184x178 & 4x3x3 & 21020.7 & 86484.7 & 58.6322 \\
	4tmn & 9400 & 213x184x178 & 4x3x3 & 21224.8 & 85989 & 58.9291 \\
	3r88 & 9948 & 203x241x214 & 4x4x4 & 23584.2 & 90457.6 & 81.9614 \\
	4gkm & 10018 & 234x134x241 & 4x3x4 & 23776.8 & 90996.8 & 65.9909 \\
	4owm & 10046 & 217x157x239 & 4x3x4 & 23757.2 & 90913.7 & 92.6124 \\
	2w66 & 10441 & 198x259x184 & 4x5x3 & 24986.9 & 93973 & 111.381 \\
	2vvn & 10458 & 199x259x185 & 4x5x3 & 24967.8 & 94680.3 & 72.0086 \\
	3ge7 & 11112 & 212x130x232 & 4x3x4 & 24214.1 & 100795 & 68.4804 \\
	3gc5 & 11402 & 210x131x233 & 4x3x4 & 24774 & 102959 & 65.0512 \\
	3rr4 & 11416 & 210x132x231 & 4x3x4 & 24687.2 & 102504 & 94.0242 \\
	3g2n & 13194 & 212x195x213 & 4x4x4 & 29288.2 & 122132 & 114.925 \\
	2wvt & 14130 & 181x167x298 & 3x3x5 & 32575.7 & 128126 & 86.4621 \\
	2wbg & 14288 & 253x208x176 & 4x4x3 & 28560.6 & 129288 & 118.505 \\
	3zdg & 15709 & 206x200x214 & 4x4x4 & 38866 & 144437 & 85.0419 \\
	3u8n & 15960 & 189x211x224 & 3x4x4 & 40327.3 & 147464 & 88.1512 \\
	3u8k & 16004 & 178x208x212 & 3x4x4 & 41707.7 & 149056 & 87.3808 \\
	2xys & 16045 & 233x229x215 & 4x4x4 & 40437 & 146928 & 100.344 \\
	1ps3 & 16157 & 205x213x246 & 4x4x4 & 33802.4 & 148038 & 94.0591 \\
	3dx1 & 16185 & 204x212x251 & 4x4x4 & 33359.3 & 147934 & 152.277 \\
	3d4z & 16185 & 205x211x248 & 4x4x4 & 33016.8 & 147051 & 100.968 \\
	3dx2 & 16185 & 206x215x246 & 4x4x4 & 33246.5 & 147399 & 106.006 \\
	3ejr & 16185 & 205x213x248 & 4x4x4 & 33829.1 & 148567 & 106.539 \\
	3f3a & 16252 & 232x181x258 & 4x3x5 & 34144.4 & 144819 & 151.338 \\
	3f3c & 16364 & 233x194x252 & 4x4x4 & 34288.8 & 147783 & 104.871 \\
	3f3d & 16364 & 235x194x254 & 4x4x5 & 34513.1 & 149136 & 104.439 \\
	3f3e & 16364 & 233x193x253 & 4x4x4 & 34231.4 & 147296 & 102.979 \\
	2wn9 & 16453 & 209x212x199 & 4x4x4 & 42401.8 & 150607 & 92.4032 \\
	4qac & 16464 & 197x215x214 & 4x4x4 & 41359.3 & 151072 & 90.1083 \\
	2wnc & 16628 & 231x210x223 & 4x4x4 & 42586.5 & 153057 & 99.1218 \\
	2x00 & 16661 & 212x214x201 & 4x4x4 & 43256.1 & 154206 & 90.0651 \\
	1e66 & 16692 & 288x256x222 & 5x5x4 & 36312.4 & 153168 & 124.091 \\
	2xj7 & 20502 & 207x307x309 & 4x5x5 & 48182.4 & 186415 & 146.431 \\
	3n7a & 25763 & 233x248x260 & 4x4x5 & 54932.7 & 236584 & 211.186 \\
	4ciw & 25848 & 266x266x266 & 5x5x5 & 55675.4 & 236737 & 164.345 \\
	2xb8 & 25848 & 263x263x263 & 5x5x5 & 56332.5 & 236229 & 160.274 \\
	3n86 & 25889 & 233x239x237 & 4x4x4 & 54431 & 236562 & 126.98 \\
	3syr & 26110 & 217x217x311 & 4x4x5 & 56471.3 & 243244 & 146.197 \\
	3l7b & 26188 & 218x218x312 & 4x4x5 & 56457.9 & 244430 & 149.303 \\
	4eky & 26216 & 217x217x311 & 4x4x5 & 56276.4 & 244775 & 217.654 \\
	3ebp & 26216 & 216x216x311 & 4x4x5 & 55922.4 & 243389 & 143.218 \\
	3n76 & 26220 & 264x264x264 & 5x5x5 & 54813.3 & 237610 & 161.143 \\
	2ymd & 32988 & 357x324x204 & 6x6x4 & 79086.5 & 303394 & 314.233 \\
	\hline
	\caption{Surface generation results of the PDBbing v2016 core set {sorted by  the number of atoms}.}
	\label{apdx:list}
\end{longtable}
	
\end{appendices}

\end{document}